\documentclass[fleqn,usenatbib]{mnras}
\usepackage{Preamble}

\usepackage{newtxtext,newtxmath}

\usepackage[T1]{fontenc}

\DeclareRobustCommand{\VAN}[3]{#2}
\let\VANthebibliography\thebibliography
\def\thebibliography{\DeclareRobustCommand{\VAN}[3]{##3}\VANthebibliography}

\usepackage{graphicx}	
\usepackage{amsmath}	
\usepackage{multirow}
\usepackage{multicol}
\usepackage{ulem}

\title[Radio variability of 3C\,279]{Multi-band Cross-correlated Radio Variability of the Blazar 3C\,279}  

\author[K. Mohana A et al.]{Krishna Mohana A$^{1}$\thanks{E-mail: krishnamohana.mon@gmail.com (KMA)}\orcidlink{0000-0002-2665-0680},
Alok C. Gupta$^{2,1}$\thanks{E-mail: acgupta30@gmail.com (ACG)}\orcidlink{0000-0002-9331-4388},
Alan P. Marscher$^{3}$\thanks{E-mail: marscher@bu.edu (APM)}\orcidlink{0000-0001-7396-3332},
Yulia V. Sotnikova$^{4,5}$\orcidlink{0000-0001-9172-7237},
S. G. Jorstad$^{3,6}$\orcidlink{0000-0001-6158-1708},
\newauthor
Paul J. Wiita$^{7}$\orcidlink{0000-0002-1029-3746},
Lang Cui$^{8}$\orcidlink{0000-0003-0721-5509},
Margo F. Aller$^{9}$\orcidlink{0000-0003-2483-2103},
Hugh D. Aller$^{9}$\orcidlink{0000-0003-1945-1840},
Yu. A. Kovalev$^{10}$\orcidlink{0000-0002-8017-5665},
Y. Y. Kovalev$^{11,10,12}$\orcidlink{0000-0001-9303-3263},
\newauthor
Xiang Liu$^{8}$\orcidlink{0000-0001-9815-2579},
T. V. Mufakharov$^{4,5}$\orcidlink{0000-0001-9984-127X},
A. V. Popkov$^{12,10}$\orcidlink{0000-0002-0739-700X},
M. G. Mingaliev$^{4,5,13}$\orcidlink{0000-0001-8585-1186},
A. K. Erkenov$^{4}$\orcidlink{0000-0002-6086-9299},
\newauthor
N. A. Nizhelsky$^{4}$\orcidlink{0000-0002-6419-7223},
P. G. Tsybulev$^{4}$\orcidlink{0000-0001-5600-8018},
Wei Zhao$^{14,15}$\orcidlink{0000-0003-4478-2887},
Z. R. Weaver$^{3}$\orcidlink{0000-0001-6314-0690}, 
D. A. Morozova$^{6}$\orcidlink{0000-0002-9407-7804} 
\\
\\
$^{1}$Aryabhatta Research Institute of Observational Sciences (ARIES), Manora Peak, Nainital 263001, India\\
$^{2}$Key Laboratory for Research in Galaxies and Cosmology, Shanghai Astronomical Observatory, Chinese Academy of Sciences, Shanghai 200030, China\\
$^{3}$Institute for Astrophysical Research, Boston University, 725 Commonwealth Avenue, Boston, MA 02215, USA \\
$^{4}$Special Astrophysical Observatory of RAS, Nizhny Arkhyz, 369167, Russia \\
$^{5}$Kazan Federal University, 18 Kremlyovskaya St, Kazan 420008, Russia \\
$^{6}$St. Petersburg State University, 7/9 Universsitetsjaya nab., St. Petersburg, 199034 Russia \\
$^{7}$Department of Physics, The College of New Jersey, 2000 Pennington Rd., Ewing, NJ 08628-0718, USA \\
$^{8}$Xinjiang Astronomical Observatory, Chinese Academy of Sciences, 150 Science 1-Street, Urumqi 830011, China \\
$^{9}$Department of Astronomy, University of Michigan, 1085 S. University Avenue, Ann Arbor, Michigan 48109, USA \\
$^{10}$Astro Space Center, Lebedev Physical Institute, Russian Academy of Sciences, Moscow, 117991 Russia \\
$^{11}$Max-Planck-Institut f$\ddot{u}$r Radioastronomie, Auf dem H$\ddot{u}$gel 69, D-53121 Bonn, Germany \\
$^{12}$Moscow Institute of Physics and Technology, Institutsky per. 9, Dolgoprudny 141700 Russia \\
$^{13}$Institute of Applied Astronomy, Russian Academy of Sciences, St. Petersburg 191187, Russia \\
$^{14}$Shanghai Astronomical Observatory, Chinese Academy of Sciences, Shanghai 200030, China \\
$^{15}$Key Laboratory of Radio Astronomy, Chinese Academy of Sciences, 210008 Nanjing, China 
}
\date{Accepted 2023 XXX. Received 2023 YYY; in original form 2023 ZZZ}

\pubyear{2023}

\begin{document}
\label{firstpage}
\pagerange{\pageref{firstpage}--\pageref{lastpage}}
\maketitle

\begin{abstract}
We present the results of our study of cross-correlations between long-term multi-band observations of the radio variability of the blazar 3C\,279. More than a decade (2008$-$2022) of radio data were collected at seven different frequencies ranging from 2 GHz to 230 GHz. The multi-band radio light curves show variations in flux, with the prominent flare  features appearing first at higher-frequency and later in lower-frequency bands. 
This behavior is quantified by cross-correlation analysis, which finds that the emission at lower-frequency bands lags that at higher-frequency bands. Lag versus frequency plots are well fit by straight lines with negative slope, typically $\sim -30$ day GHz$^{-1}$. We discuss these flux variations in conjunction with the evolution of bright moving knots seen in multi-epoch VLBA maps to suggest possible physical changes in the jet that can explain the observational results. Some of the variations are consistent with the predictions of shock models, while others are better explained by a changing Doppler beaming factor as the knot trajectory bends slightly, given a small viewing angle to the jet.
 
\end{abstract}




\section{Introduction}
The blazar class of active galactic nuclei (AGNs) exhibits
multi-wavelength (MW) emission dominated by non-thermal radiation originating in a relativistic jet pointing towards the Earth, which causes the radiation to be strongly Doppler boosted \citep{urry1995}.
The emission from blazars is highly variable over time scales ranging from minutes to days 
to months to years \citep[e.g.,][and references therein]{mattox1997,
hartman2001,boettcher2004,foschini2013,orienti2014,abdo2015,boettchergalaxy2019}.
From the observational perspective, blazars are sub-classified as flat-spectrum radio quasars (FSRQs) and BL Lacertae objects (BL Lacs).  
The spectrum of FSRQs contain broad emission lines with equivalent widths 
(EW) \textgreater $5$ \AA, ~whereas BL Lacs have EW \textless $5$ \AA ~\citep{stocke1991,stickel1991}.
According to the classification proposed by  \citet{ghisellini2009a}, 
FSRQs have greater broad-line region (BLR) luminosities ($L_{\mbox{BLR}}/L_{\mbox{Edd}} >5\times10^{-4}$) than BL Lacs, where $L_{\mbox{Edd}}$ is the Eddington luminosity for the supermassive black hole at the center of the AGN. \\
\\
The broad-band spectral energy distribution (SED) of blazars 
spans the radio to $\gamma$-ray bands and contains a double-hump structure \citep[e.g.][]{fossati1998}. 
The matter composition inside the blazar jets and the dominant particle 
population responsible for the observed emission in both humps is still not certain \citep{boettchergalaxy2019}. 
However, most blazar studies consider the particles to include ultrarelativistic electrons and either positrons (leptonic scenario) or
protons (hadronic scenario) to be the primary emitters 
\citep{bottcher2013ApJ}. 
In either scenario, the first hump of the SED, 
which peaks between the infrared (IR) and X-ray bands is explained by synchrotron 
emission \citep{blumenthal1970RvMP} originating from leptons in the relativistic jet (\citealt{abdo2010b,bottcher2013ApJ,boettchergalaxy2019}, and references therein). \\
\\
One of the most effective ways to understand jet emission and its dynamics is to study the changes in physical properties of blazars required to produce multi-frequency 
light-curve variations. 
In this regard, cross-correlation techniques can be an efficient method 
to extract the temporal trends of observed variability 
and to identify plausible connections  
across various blazar emission bands (e.g., \citealt{liodakis2018}, and references therein). It has been noticed from such multi-band observations that the variability patterns in different wavebands are sometimes
correlated, but often not (e.g., \citealt{hayashida2012ApJ,boettchergalaxy2019}, and  references therein). Understanding the physical origin of the variability behaviour of blazars is one of the major open questions in the field of AGN research. Multi-band cross-correlation studies help in constraining the radiative mechanism and the location of the emission region, and in understanding whether or not the emission processes at various energy bands are connected. \\
\\
3C\,279 is a FSRQ ($z = 0.5362)$ \citep{marziani1996ApJS} containing a black hole with mass in the range of (3--8) $\times \ 10^{8}M_\odot$ \citep{Gu2001,woo2002}. The presence of apparent superluminal motion of radio knots in this source was first reported by \citet{cohen1971}. The radio morphology of 3C\,279 exhibits a bright stationary compact core and a thin jet then consisting of six knots aligned with position angle (PA) $\sim205^{\circ}$ and extending out $4.7''$ \citep{1983ApJ...273...64D}. The source has often been observed with high resolution very long baseline interferometry (VLBI), which reveals the presence of a single-sided jet extending southwest on parsec scales with bright knots expelled from the core region. The features in the jet display various projected apparent speeds and polarization angles \citep[e.g.,][]{unwin1989ApJ,wehrle2001ApJS,jorstad2005AJ,chatterjee2008ApJ}. Polarimetric studies using observations from the Very Long Baseline Array (VLBA) at 43 GHz have found the electric field vector usually to be oriented along the jet direction on pc to kpc scales, indicating that the magnetic field is mainly perpendicular to the relativistic jet \citep{jorstad2005AJ}. During the last decade, several intensive MW studies of 3C\,279 have examined cross-correlated variability in an effort to understand the emission mechanism in this FSRQ.  
Such intensive studies have shown that the broad-band emission of 3C\,279 displays complex behaviour, with variability ranging over time scales from minutes to years across the electromagnetic bands (e.g., \citealt{hayashida2012ApJ,fuhrmann2014,pati2017FrASS,prince2020ApJ,rajput2020MNRAS}, and references therein).\\
\\
The prime motivation of this work is to search for correlations between the different radio frequency emissions of 3C\,279 using long-term observations. A few earlier studies have discussed multi-frequency radio cross-correlations of 3C\,279 \citep{wang2008,deng2008,yuan2012}. However, they included observations covering a much shorter or non-overlapping time of observations and also less dense coverage of frequency bands than presented in this work. These studies suggested that emission at lower-frequency bands lags that of the higher-frequency ones. However, it is vital to check the significance of such a pattern  with more extended long-term observations comprising abundant data covering a broad range of frequencies. The presence of positive (or negative) correlations between the different radio emission bands could significantly constrain the origin of the radio variability and identify plausible physical scenarios. For example, this study can investigate whether the radio variability in this blazar is always consequence of a shock wave propagating down an adiabatic, conical, relativistic jet \citep[e.g.,][]{marscher1985,hughes1989ApJ}.  
In addition, we can address the importance of changes in synchrotron self-absorption (SSA) opacity between the higher and lower frequency emission \citep{aller1985ApJS}. Furthermore, the study can allow an assessment of the extent to which bends in the jet flow affect the flux density as the Doppler factor changes.
\\
\\
To understand the physical processes responsible for the observed blazar emission, it is important to
observe the changes in the radio jet and relate them to multi-wavelength variability (e.g., \citealt{fuhrmann2014,weaver2022ApJS}, and references therein).
In this work, we present a long-term ($\sim$14 years) multi-band radio cross-correlation study of 3C\,279. Our study includes radio data ranging from 2.25 GHz to 230 GHz, with observations spanning from  August 2008 to October 2022. This is first multi-band radio cross-correlation study on 3C\,279 using more than a decade of observations made at seven different radio bands. \\
\\
In our study, the data were collected from six radio observatories, which include unpublished measurements we made at the RATAN-600 and the Xinjiang Astronomical Observatory--Nanshan station radio telescopes (XAO-NSRT) along with archival data, the bulk of which were taken by our groups. 
These observations were made at 7 different frequency bands. Details of the radio data and the analysis are
described in Section~\ref{sec:data_radio}. In Section~\ref{sec:results} and Section~\ref{sec:discussion} we present and discuss the results, respectively. Our conclusions are provided in Section~\ref{sec:conclusions}.

\section{Observations and data analysis}\label{sec:data_radio}
\subsection{Radio telescopes and observations}
This long-term ($\sim$14 years, mainly 4 August 2008 to 31 October 2022) radio study, ranging from 2 GHz to 230 GHz, was made possible by the accumulation of observations from various radio observatories. 
Information on the radio telescopes and corresponding data coverage periods used in this work are listed in Table~\ref{tab:radio_data_info}. \\

\begin{table}
{  \centering
  \caption{Radio observatory data used in this work}
  \label{tab:radio_data_info}
\begin{tabular}{ccc}
  \hline
Observatory  & Data coverage time  & Bands (GHz) \\
\hline
RATAN-600   & 04-08-2008 to 31-10-2022  & 2.25, 4.7, 8.2 \\
            & 04-08-2008 to 31-10-2022   & 11.2, 22.3 \\
Effelsberg  & 04-08-2008 to 01-01-2015  & 2.64, 4.8, 8.35 \\
&    04-08-2008 to 01-01-2015  & 10.45,  21.7\\
XAO-NSRT    & 29-03-2017 to 29-10-2022 &4.8 \\
            & 24-08-2018 to 17-10-2022   & 23.6 \\
UMRAO        & 	18-11-2009  to 	28-04-2012 &  4.8\\
            & 07-09-2009 to 16-05-2012 &  8.0\\
VLBA       & 04-08-2008 to 24-06-2022 &  43 \\
SMA        & 04-08-2008 to  09-09-2022 &  230\\
\hline 
  \end{tabular}\\
\noindent\\
}
{\footnotesize{Note: The mean cadence (in days) of the light curves after merging 
the observations from different observatories in order to construct a complete light curve in each radio band
are: 9 days for 230 GHz; 35 days for 43 GHz; 23 days for 23 GHz; 28 days for 11 GHz; 20 days for 8 GHz; 16 days for 4.8 GHz
and 43 days for 2.5 GHz.}}
 \end{table}

\noindent 
The data include both publicly available archival and published observations, as well as data reduced from raw observations made at various radio telescopes. The archival data includes observations from: the Fermi-GST AGN Multi-frequency Monitoring Alliance (F-GAMMA), made at the Effelsberg telescope in Germany, with radio fluxes published in \citet{angelakis2019}; the University of Michigan Radio Astronomy Observatory\footnote{\url{https://dept.astro.lsa.umich.edu/datasets/umrao.php}}
\citep[UMRAO;][]{aller1985}; the Submillimeter Array\footnote{\url{http://sma1.sma.hawaii.edu/callist/callist.html}} \citep[SMA;][]{gurwell2007} in Hawaii; and the VLBA by the Boston University (BU) 
group: the VLBA–BU–BLAZAR programme\footnote{\url{http://www.bu.edu/blazars/BEAM-ME.html}}.
Here we have utilized the relatively recent observations of 3C\,279 made with
XAO-NSRT, Urumqi, China and the very extensive data from the RATAN-600 observatory in Zelenchukskaya, Russia. 
3C\,279 was part of multi-frequency radio monitoring campaigns at these observatories. \\
\\
The XAO-NSRT is operated by the Urumqi observatory,  which uses the 26-m parabolic antenna of the Nanshan radio telescope, China ~(\citealt{sun2006,marchili2010}, and references therein). This facility consists of a single-beam dual-polarization receiver, constructed by the Max-Planck-Institut f\"ur Radioastronomie (MPIfR), with central frequency of 4.80 GHz and bandwidth of 600 MHz. The flux density was measured in cross-scan mode, with each scan comprised of eight sub-scans (four in azimuth and four in elevation) over the source position. The data were acquired in this fashion at C-band (4.8 GHz) and K-band (23.6 GHz). The XAO-NSRT data used in this work were reduced by following the data calibration and reduction procedures as explained in Section~2 of \citet{marchili2011};  see also \citet{marchili2012} and \citet{liu2015}.

\begin{table}
\caption{RATAN-600 radiometer parameters: central frequency $f_0$, bandwidth $\Delta f_0$, detection limit for sources per transit $\Delta F$. ${\rm FWHM}_{\rm {RA} \times \rm {DEC}}$ is the average angular resolution along RA and Dec.} 

\label{tab:rad}
\centering
\begin{tabular}{cccc}
\hline
$f_{0}$ & $\Delta f_{0}$ & $\Delta F$ &  {FWHM$_{\rm{RA} \times \rm{Dec}}$}\\
GHz   &   GHz    &  mJy/beam   &     \\
\hline
$22.3$ & $2.5$  &  $50$ & $0\farcm14 \times 0\farcm80$  \\
$11.2$ & $1.4$  &  $15$ & $0\farcm28 \times 1\farcm64$ \\ 
$8.2$  & $1.0$  &  $10$ & $0\farcm39 \times 2\farcm43$   \\ 
$4.7$  & $0.6$  &  $8$  & $0\farcm68 \times 4\farcm02$   \\ 
$2.25$  & $0.08$  &  $40$ & $1\farcm42 \times 8\farcm50$  \\ 
$1.25$  & $0.08$ &  $200$ & $2\farcm60 \times 16\farcm30$ \\
\hline
\end{tabular}
\end{table} 

\noindent
The radio telescope RATAN-600 \citep{1993IAPM...35....7P} has a 600-m circular multi-element antenna that provides measurements of 1--22 GHz broad-band spectra simultaneously (within 1--2 min) when a source moves along the focal line where the receivers are located. 
The angular resolution depends on the antenna elevation angle, and its values along RA ($\rm FWHM_{RA}$) and Dec ($\rm FWHM_{Dec}$) calculated for an average Dec value ($\delta\sim30^{\circ}$) and are presented in Table~\ref{tab:rad}. 
The observations were carried out with six-frequency radiometers at 1.25 (0.96 earlier), 2.25, 4.7 (3.9 earlier), 8.2, 11.2, and 22.3 GHz.  The observations of 3C\,279 and astronomical flux density calibrators were made at their upper culminations. The objects were scanned with a fixed antenna throughout the daily rotation of the Earth and the horizontal location of all input horns in the frequency bands  from 1 to 22 GHz in the focal plane of the antenna. We used the following six flux density secondary calibrators: 3C 48, 3C 138, 3C 147, 3C 161, 3C 286, and NGC 7027. The flux density values were calculated based on the scales provided by \citet{baars1977} and \citet{2013ApJS..204...19P,2017ApJS..230....7P}. The measurements of the calibrators were corrected for linear polarization and angular size according to the data from \citet{1994A&A...284..331O} and \citet{1980A&AS...39..379T}.\\
\\
The RATAN-600 measurements were processed using the automated data reduction system \citep{1999A&AS..139..545K,2011AstBu..66..109T,2016AstBu..71..496U,2018AstBu..73..494T} and the Flexible Astronomical Data Processing System (\textsc{FADPS}) standard package modules \citep{1997ASPC..125...46V} for the broad-band RATAN continuum radiometers. Some of the RATAN-600 measurements are  presented in the on-line catalogue BLcat\footnote{\url{https://www.sao.ru/blcat/}}. It should be noted that all RATAN-600 data, except those at 1.25 GHz, were used in the present work. The data at 1.25 GHz were not included due to sparser  coverage at this frequency band, which made them inadequate for detailed comparisons with other bands.

\subsection{Very Long Baseline Array Observations}\label{sec:vlba}
We use VLBA observations of 3C~279 performed within the VLBA-BU-BLAZAR programme and its successor Blazar Entering Astrophysical Multi-Messenger Era (BEAM-ME)  programme of monitoring of a sample of $\gamma$-ray blazars which cover the period from 2008 August to 2022 July. The observations were carried out in continuum mode, with both left (L) and right (R) circular polarization signals recorded at a central frequency of 43.10349~GHz using four intermediate frequency bands (IFs), each of 64~MHz width over 16 channels. The monitoring had a roughly monthly cadence of 24~hrs duration at each epoch for either 32 or 33 sources of the sample, with 3C~279 observed at each epoch with 7--9 scans (5 min per scan). Except for several epochs, all 10 VLBA antennas participated in the observations.\\
\\
The data were correlated at the National Radio Astronomy Observatory (NRAO; Soccoro, NM) and reduced using the {\it Astronomical Image Processing System} software (AIPS) provided by NRAO, with the latest version corresponding to the epoch of observation, and {\it Difmap} \citep{Shepherd1997} in the manner described by \cite{Jorstad2017}.
The average size of a synthetic beam for 3C~279 is 0.36$\times$0.15~mas$^2$ at position angle PA=$-$10~deg. To construct the light curve at 43~GHz from the VLBA data flux densities are integrated over images. We perform twice per year observations with Jansky Very Large Array (JVLA) at 43~GHz of several compact sources from the sample, close to corresponding VLBA epochs, to check the amplitude calibration, and employ also measurements from the Mets\"ahovi Radio Observatory (Aalto Univ., Finland) blazar monitoring programme at 37~GHz, which includes many sources from our sample. For 3C~279 the integrated flux density is usually within 10 percent of the corresponding JVLA measurement. In addition, this study uses results of the 3C~279 parsec-scale jet kinematics over the period from 2008 August to 2018 December based on these data and published in \cite{weaver2022ApJS}. 

\subsection{Cross-correlation analysis}\label{sec:CCF_method}
We carried out cross-correlation between different radio light curves using the z-transformed discrete correlation function (ZDCF) method \citep{alexander1997,alexander2013}.
This method is applicable to both evenly and unevenly sampled data, but is particularly appropriate when the observed data are unevenly sampled and sparse. The ZDCF binning algorithm adopts the same idea as that of the discrete correlation
function (DCF) developed by \citet{edelson1988}. However, in ZDCF, the statistical significance for each bin is made sufficiently high by varying that bin's width.
Consider that there are $n$ pairs of flux densities \{$a_{i}$,$b_{i}$\} at a particular time-lag bin. 
Then the cross-correlation function (CCF)
at the lag $\tau$ is calculated using the correlation coefficient \citep{alexander2013}
\begin{equation}
	r=\dfrac{\sum_{i}^{n}(a_{i}-\overline{a})(b_{i}-\overline{b})/(n-1)}{S_{a}S_{b}},
\end{equation}
where $\overline{a}$, $\overline{b}$ represent the mean bin values and $S_{a}$, $S_{b}$ correspond to the 
standard deviations,
\begin{equation}
	S_{a}^{2}=\dfrac{1}{n-1}\sum_{i}^{n}(a_{i}-\overline{a})^{2};
\end{equation}
$S_{b}^2$ is defined similarly. However, such usage of the sample variance $S_{r}$ could be inaccurate due to the 
high skewness of the sample distribution of $r$ \citep{alexander2013}.
If $a$ and $b$ are drawn from a bivariate normal distribution, then one can 
convert $r$ to an approximately normally distributed random variable,
Fisher's $z$ (\citealt{alexander2013}, and the references therein). Defining

\begin{equation}
	z\equiv\dfrac{1}{2}\log\left(\dfrac{1+r}{1-r}\right),~\zeta\equiv\dfrac{1}{2}\log\left(\dfrac{1+\rho}{1-\rho}\right),~r\equiv\mbox{tanh}z,
\end{equation}
where $\rho$ is the unknown population correlation coefficient of the bin. 
Transforming back to $r$, the error interval ($\pm1\sigma$) is calculated as
\begin{equation}
	\delta r_{\pm}=\left|\mbox{tanh}(\overline{z}(r)\pm S_{z}(r))-r\right|,
\end{equation}
where $S_{z}(r)$ is the variance of $z$. In the ZDCF, equal population binning is considered rather than binning by equal  $\Delta\tau$.
Obtaining  convergence for $z$-transform requires the minimum number of points  per bin ($n_{\rm{min}}$) to be 11.
Also, to avoid bias, the dependent pairs (e.g., $\tau_{i,j}$, $\tau_{i,k}$ for the case of estimation of time difference in $\tau_{i,j}$) are dropped. 
We made use of the openly available FORTRAN 95 routine\footnote{\url{https://www.weizmann.ac.il/particle/tal/research-activities/software}} for the specific procedure explained in \citet{alexander1997}
to carry out the cross-correlation. While performing the ZDCF analysis, we chose the
default $n_{\rm{min}} = 11$ for each bin. We used 1000
Monte Carlo runs for error estimation of the coefficients.
Also, we did not consider the points for which the lag was zero.\\
\\
To estimate the significance of a DCF peak, we followed the method described in \citet{maxmoerbeck2014}. For this, we simulated 
a total of 1000 light curves at each frequency band with the same power spectral density and the flux distribution function of the original 
light curve using the algorithm provided by \cite{emmanoulopoulos2013}, as realized by the DELightcurveSimulation code \citep{connolly2015}\footnote{\url{https://github.com/samconnolly/DELightcurveSimulation}}.
Further, from the distribution of the DCF between simulated light
curve pairs (using the same method as for the real data), the thresholds for $1\sigma$, $2\sigma$ and $3\sigma$ significance were calculated at each lag. We only consider the significant DCF peak closer
to zero lag and discard any peaks near the edges of the
temporal span considered. The peaks present at large values of $\tau$ might be due to the several
smaller fluctuations present in the light curves \citep[e.g.,][]{meyer2019ApJ,maxmoerbeck2014}. 
Also, we notice that the overlap between the light curves is smaller for larger values
of $\tau$, hence we consider these peaks less credible.
After identifying a significant DCF peak, to estimate the exact location of cross-correlation function peak with corresponding uncertainties, we implement the maximum likelihood method of \citet{alexander2013}
by using the openly available FORTRAN 95 routine ({\sc PLIKE})\footnote{\url{https://www.weizmann.ac.il/particle/tal/research-activities/software}}. We note that this method estimates a fiducial interval rather than the traditional confidence interval. The approach taken here is similar to Bayesian statistics, where the normalized likelihood function (i.e. fiducial distribution) is interpreted as expressing the degree of belief in the estimated parameter, and the 68 percent interval around the likelihood function’s maximum represents the fiducial interval.
Here a positive lag for a DCF tagged as `light curve 1 versus light curve 2’ means that the emission in light curve 2 lags that  of light curve 1, while a negative lag means 2 leads 1. 

\section{Results}\label{sec:results}
\subsection{Decade-long multi-band radio light curve of 3C\,279}
\begin{figure*}
	\centering
\includegraphics[scale=0.38]{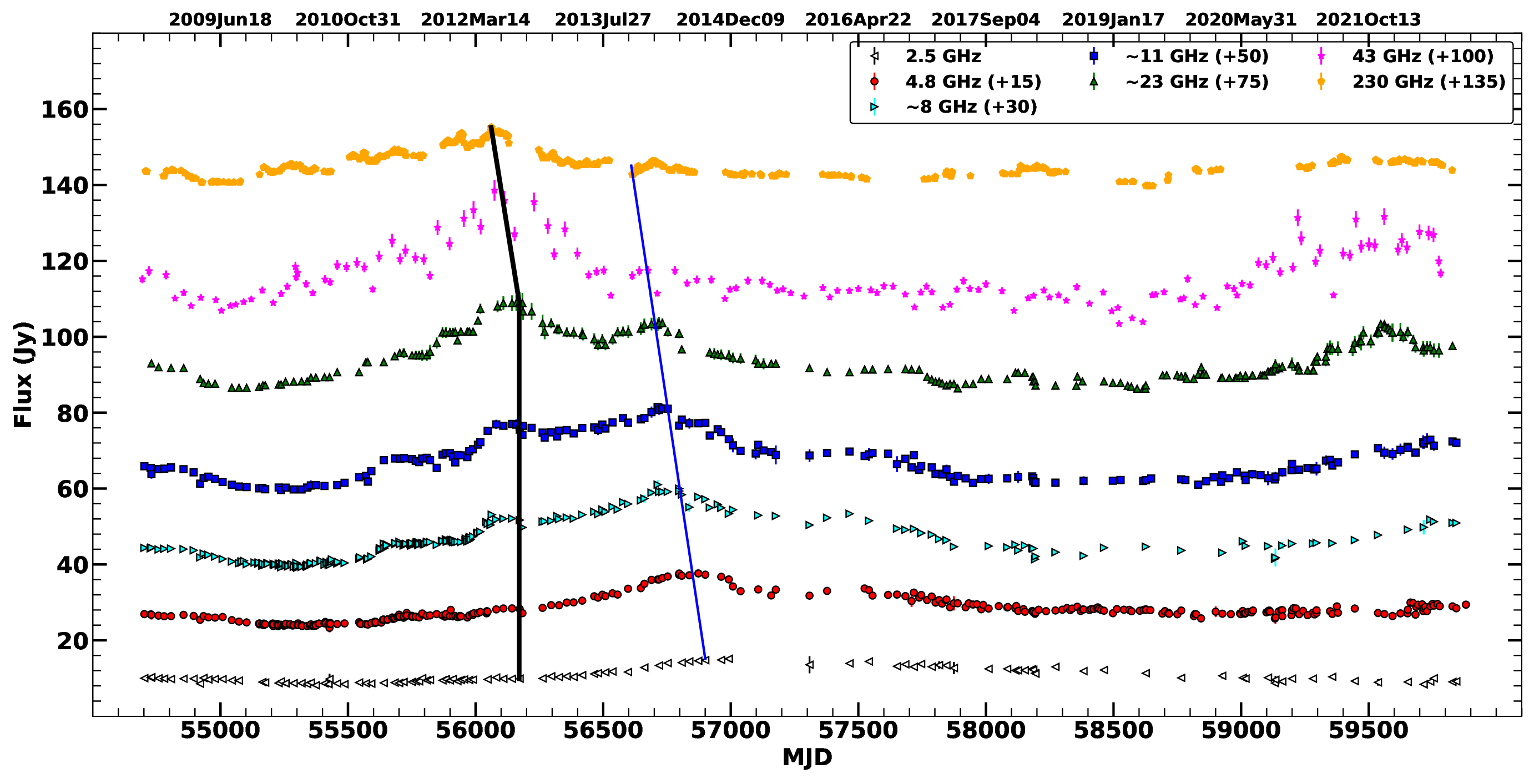}
	\caption{Multi-band radio lightcurve of 3C\,279. For clearer visual presentation, the individual light curves are shown with offsets, which are provided in the inset label. The solid black and thin blue slanted lines connect the peaks of the first and second outbursts, respectively.}
    \label{fig:multiband_radio_lightcurve}
\end{figure*}

Using the more than decade-long radio observations from various observatories, we have created a multi-band radio light curve of 3C\,279 (Fig.~\ref{fig:multiband_radio_lightcurve}).

We have merged the observations from different observatories in order to construct a complete light curve in each radio band. In Figure~\ref{fig:multiband_radio_lightcurve}, the light curve at 2.5 GHz is the result of merging the observations from RATAN-600 at 2.25 GHz and F-GAMMA at 2.64 GHz; the light curve at 4.8 GHz  combines data from RATAN-600 at 4.7 GHz, F-GAMMA at 4.8 GHz, XAO-NSRT at 4.8 GHz, \& UMRAO at 4.8 GHz; the light curve at 8 GHz comes from merging observations from RATAN-600 at 8.2 GHz, F-GAMMA at 8.35 GHz, and UMRAO at 8.0 GHz; the light curve at 11 GHz combines the observations from RATAN-600 at 11.2 GHz \& F-GAMMA at 10.45 GHz; finally, the light curve at 23 GHz combines our observations from RATAN-600 at 22.3 GHz and XAO-NSRT at 23.6 GHz with those of F-GAMMA at 21.7 GHz. The long term light curves data at 43 GHz and 230 GHz are taken from VLBA and SMA, respectively.
The mean cadence of these merged light curves at different frequency bands is provided in Table~\ref{tab:radio_data_info}.
We note that outliers/very short-term fluctuations are present in the original light curves at 23 and 230 GHz, which may be caused by turbulence or some other fast random process. Since these short time-scale fluctuations do not play a significant role in the ZDCF correlations, we have smoothed the original light curves at 23 and 230 GHz using the median smoothing algorithm (with data points $n=5$) available in the SPMF open-source data mining library\footnote{\url{https://www.philippe-fournier-viger.com/spmf/MedianSmoothing.php}} \citep{fournier2016}. This provides for a clearer representation of longer term variation patterns.\\
\\
From visual inspection of the multi-band radio light curves of 3C\,279 (Fig.~\ref{fig:multiband_radio_lightcurve}), we noticed that the most prominent peak in the
emission first occurs around MJD 56100 (22 June 2012) at higher radio frequencies (230 GHz and 43 GHz). The   emission amplitude typically declines with decreasing frequency. Such a trend was also reported by \citet{larinov2020MNRAS} for 3C\,279.
However, they did not report any cross-correlation study on the different radio light curves.  
Visual inspection suggests that the radio emission at lower-frequency bands  usually lags behind that at the higher-frequency bands. To quantify the lags/leads between different radio light curves, we carried out cross-correlation analyses.  

\subsection{Intra-band radio cross-correlation}
\begin{figure*}
	\centering
	\includegraphics[scale=0.38]{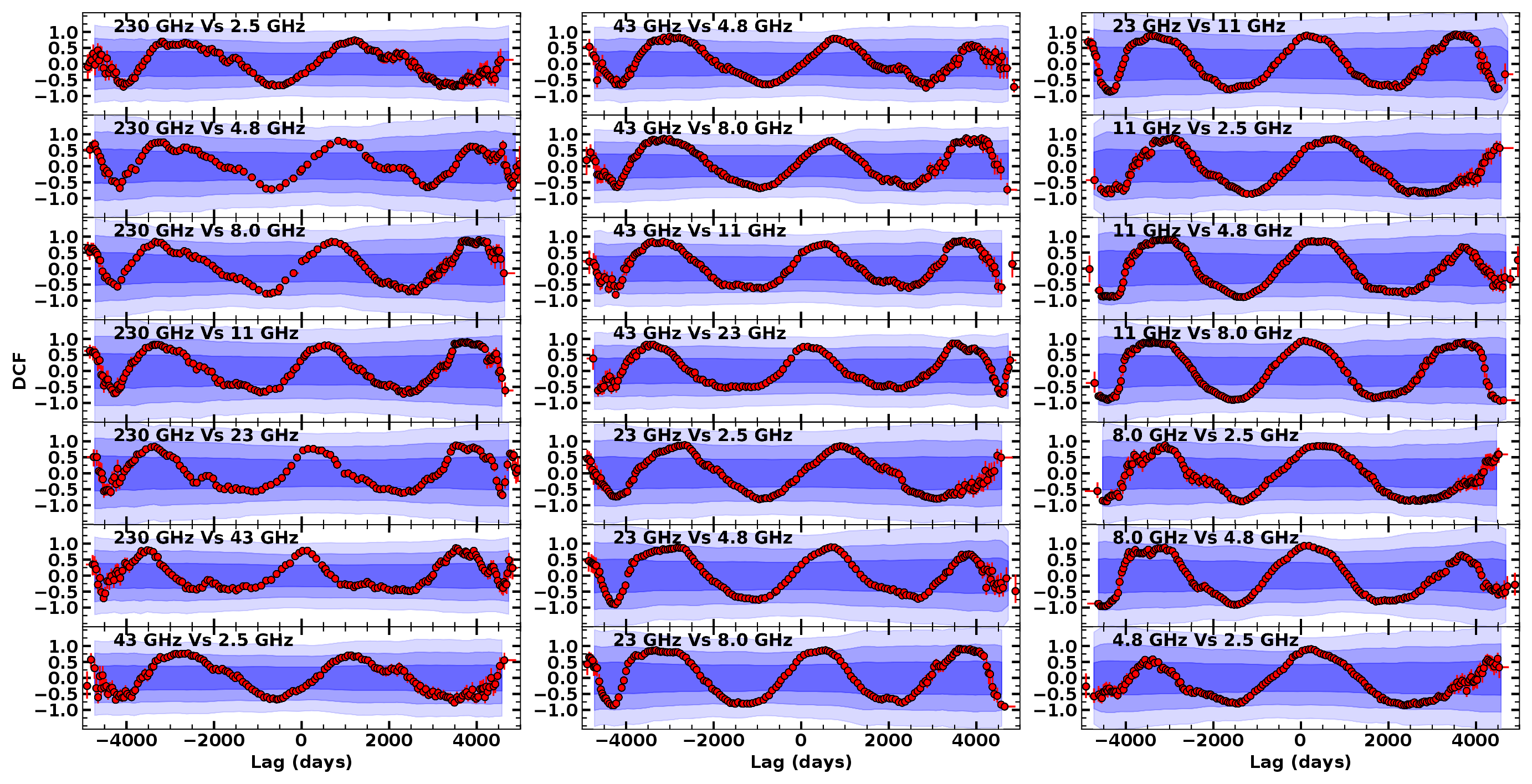}
	\caption{Cross-correlation plots between different combinations of radio bands. Here the positive lag for DCF tagged as `frequency 1 Vs frequency 2’ means light curve 2 variations lag those of light curve 1. The blue coloured contours from dark to light represent the $1\sigma$, $2\sigma$ and $3\sigma$ significance levels, respectively.}
    \label{fig:dcf_multiplots}
\end{figure*}

Adopting the ZDCF cross-correlation methodology, as explained in Section~\ref{sec:CCF_method}, we carried out intra-band cross-correlations between all possible combinations of light curves.
The cross-correlation plots (21 combinations) are shown in Figure~\ref{fig:dcf_multiplots}. The peak DCF value and the corresponding lag with $1\sigma$ error are provided in Table~\ref{tab:zdcf_results}. 
From these cross-correlation results (Fig.~\ref{fig:dcf_multiplots} and Table~\ref{tab:zdcf_results}), we find that the lower frequency light curve lags behind the higher frequency one for nearly for every pair. 
The information on lags/leads and the DCF values for all the combinations of light curves are
given in Table~\ref{tab:zdcf_results}. 

\begin{table}
  \centering
  \caption{ZDCF Results for Radio Lightcurves}
  \label{tab:zdcf_results}
\begin{tabular}{ccc}
\hline
Light curves & DCF  & Lag (days)  \\ \hline
230 GHz vs 2.5 GHz & $0.74^{+0.06}_{-0.06}$  & $+1207.0^{+81.1}_{-106.9}$\\
230 GHz vs 4.8 GHz & $0.80^{+0.03}_{-0.03}$  & $+833.8^{+98.6}_{-82.1}$\\ 
230 GHz vs 8.0 GHz & $0.84^{+0.03}_{-0.03}$  & $+682.3^{+99.2}_{-49.0}$\\
230 GHz vs 11 GHz & $0.80^{+0.04}_{-0.04}$  & $+615.0^{+58.1}_{-124.0}$\\ 
230 GHz vs 23 GHz & $0.76^{+0.03}_{-0.03}$  & $+269.3^{+79.8}_{-134.2}$\\   
230 GHz vs 43 GHz & $0.79^{+0.03}_{-0.04}$  & $+133.1^{+47.9}_{-103.3}$\\
  \hline
43 GHz vs 2.5 GHz & $0.71^{+0.06}_{-0.07}$  & $+1132.0^{+151.4}_{-99.3}$\\
43 GHz vs 4.8 GHz & $0.79^{+0.04}_{-0.04}$  & $+735.5^{+102.0}_{-51.6}$\\
43 GHz vs 8.0 GHz & $0.79^{+0.04}_{-0.04}$  & $+699.0^{+47.7}_{-92.1}$\\
43 GHz vs 11 GHz &  $0.77^{+0.04}_{-0.05}$  & $+530.1^{+86.41}_{-101.0}$\\
43 GHz vs 23 GHz &  $0.76^{+0.04}_{-0.04}$  & $+149.9^{+106.2}_{-102.7}$\\
\hline
23 GHz vs 2.5 GHz & $0.85^{+0.03}_{-0.04}$  & $+891.7^{+86.0}_{-80.4}$\\
23 GHz vs 4.8 GHz &  $0.90^{+0.02}_{-0.02}$  & $+676.7^{+75.4}_{-41.0}$\\
23 GHz vs 8.0 GHz &  $0.87^{+0.02}_{-0.03}$  & $+554.7^{+43.2}_{-96.8}$\\
23 GHz vs 11 GHz &  $0.89^{+0.02}_{-0.02}$  & $+134.9^{+56.6}_{-53.3}$\\
\hline
11 GHz vs 2.5 GHz &  $0.85^{+0.03}_{-0.04}$  & $+761.1^{+66.8}_{-130.90}$\\
11 GHz vs 4.8 GHz &  $0.86^{+0.03}_{-0.03}$  & $+529.3^{+63.5}_{-276.5}$\\
11 GHz vs 8.0 GHz &  $0.94^{+0.01}_{-0.01}$  & $+65.8^{+51.6}_{-41.3}$\\
\hline
8.0 GHz vs 2.5 GHz &  $0.85^{+0.03}_{-0.03}$  & $+375.0^{+196.1}_{-95.7}$\\
8.0 GHz vs 4.8 GHz &  $0.94^{+0.01}_{-0.01}$  & $+185.6^{+44.3}_{-103.6}$\\

\hline
4.8 GHz vs 2.5 GHz &  $0.90^{+0.02}_{-0.02}$  & $+237.0^{+46.1}_{-73.3}$\\
\hline  
  \end{tabular}
 \end{table}

\subsection{Connection between lag and frequency}
\begin{figure*}
	\centering
	\includegraphics[scale=0.40]{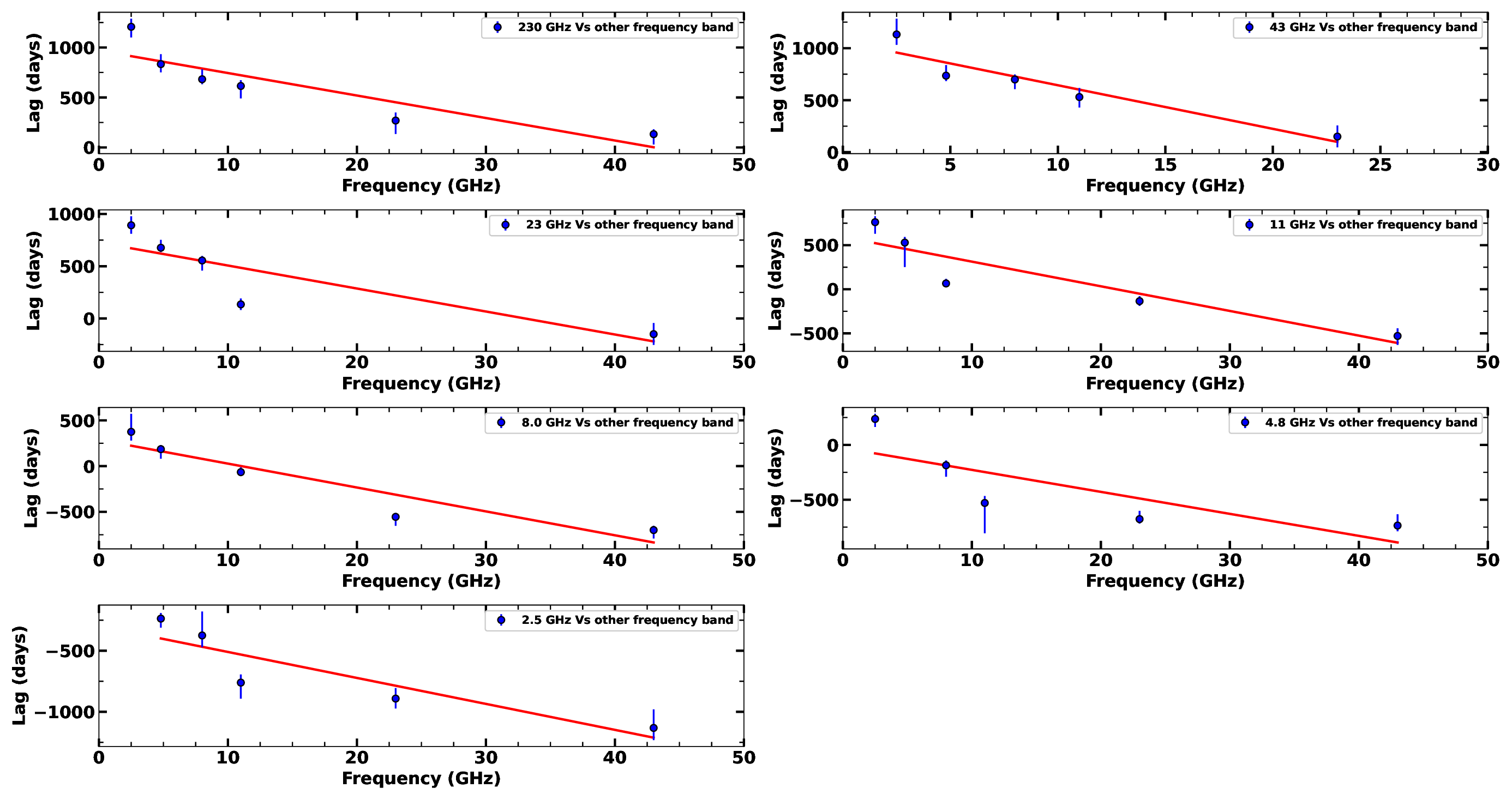}
	\caption{Lag versus frequency for the seven different radio bands. The solid red line represents the straight-line fit to the data with the best-fit parameters provided in the text (Eqns. (6) -- (12)).}
    \label{fig:freq_vs_lag_plot}
\end{figure*}

We next carried out an investigation of the connection/dependency of the observed lags with the radio emission at different frequencies. We used the estimated lag values quoted in Table~\ref{tab:zdcf_results} and plotted lags versus frequency for each radio band against the rest of the bands (Fig.~\ref{fig:freq_vs_lag_plot}).
To derive an empirical relationship between lag and frequency, we fitted a straight line defined as: 
\begin{equation}
    \tau_{\nu_{p}}~({\rm days})= [m~(\pm\sigma_{m})\times \nu~({\rm GHz}) ]+ c~(\pm\sigma_{c}),
\end{equation}
 where $\tau_{\nu_{p}}$ is the lag value of particular frequency data ($\nu_{p}$) with respect to the other frequency bands ($\nu$). The slope and intercept values are denoted by
 $m$ and $c$, along with the associated uncertainties as $\sigma_{m}$ and 
 $\sigma_{c}$, respectively. The straight-line fits are shown as  solid red lines in Figure~\ref{fig:freq_vs_lag_plot}. These relations between lag and frequency for different frequency bands are as follows:
\begin{eqnarray}
    \tau_{230\rm{GHz}} = [-22.50~(\pm5.85) \times \nu ]+ 969.57~(\pm121.62) \\
    \tau_{43\rm{GHz}}= [-41.91~(\pm8.42)\times \nu]+1062.55~(\pm102.69) \\
    \tau_{23\rm{GHz}}=[-22.00~(\pm7.35)\times \nu]+726.59~(\pm149.29) \\
    \tau_{11\rm{GHz}}=[-27.99~(\pm6.96)\times \nu]+593.31~(\pm154.81) \\
    \tau_{8.0\rm{GHz}} =[-26.12~(\pm5.63)\times \nu]+288.61~(\pm126.57) \\
    \tau_{4.8\rm{GHz}} = [-20.07~(\pm8.73)\times \nu]-26.87~(\pm197.85) \\
     \tau_{2.5\rm{GHz}}= [-21.26~(\pm6.00)\times \nu]-297.58~(\pm136.51) 
\end{eqnarray}
A negative slope was observed for every case, quantifying that the lower-frequency emissions lag behind those at higher frequencies. Five of the six better-defined slopes, i.e., Eqns. (6), (8)--(11), are consistent with $\sim -30 {\rm ~d~ GHz}^{-1}$.

\section{Discussion}\label{sec:discussion}
This work presents a systematic study of the multi-band cross-correlated radio variability of the blazar 3C\,279.
For this, we used the radio observations made over more than a decade from various radio telescopes spanning frequency bands
from 2 GHz to 230 GHz. 
\subsection{Comparison with earlier studies}
A few multi-frequency radio cross-correlation studies for 3C\,279
have been reported earlier \citep{wang2008,deng2008,yuan2012}. 
However, these studies include observations covering shorter and/or non-overlapping periods of observations; they also employ fewer frequency bands than presented in this work.
\citet{wang2008} carried out cross-correlations using radio light curves at 8 GHz (data time period: 1965 to 2000), 22 GHz (data time period: 1980 to 2005) and 37 GHz (data time period: 1980 to 2005). They found that the 8 GHz variations lag behind those at 22 GHz by $\sim69$ days, and 22 GHz data lag behind 37 GHz by $\sim33$ days.  
\citet{wang2008} suggested that the jet magnetic field primarily determines the observed time lag between bands arising from synchrotron radio emission. If the magnetic field strength is weak, then, because of the slower cooling of synchrotron radiation at lower frequencies, the observed time lag will be longer in the lower than in the higher-frequency bands. 
Such weak magnetic fields could arise either due to the location of the emitting region being far away from the central black hole or by weaker magnetic field compression by the shock waves in the emitting region.\\
\\
\citet{deng2008} carried out a cross-correlation between the Mets{\"a}hovi radio light curves at 22 GHz and 37 GHz (data time period: 1980--2005). 
Their results showed that there is a strong correlation between the two radio frequencies.   
Using the DCF technique, they found that 37 GHz light curve leads the 22 GHz  by $\sim140$ days. However, with the ZDCF method they found the 22 GHz data leading the 37 GHz by $\sim168$ days and they stated that their results were not reliable because of large uncertainties.\\
\\
\citet{yuan2012} presented cross-correlation results using light curves at 4.8 GHz, 8 GHz, and 14.5 GHz (data time period: 2009 September to 2010 July). Since the data length in this work was quite short (only $\sim$ 9 months), no significant time lags between the three frequency bands were found.\\ 

\begin{figure*}
	\centering
 \includegraphics[scale=0.38]{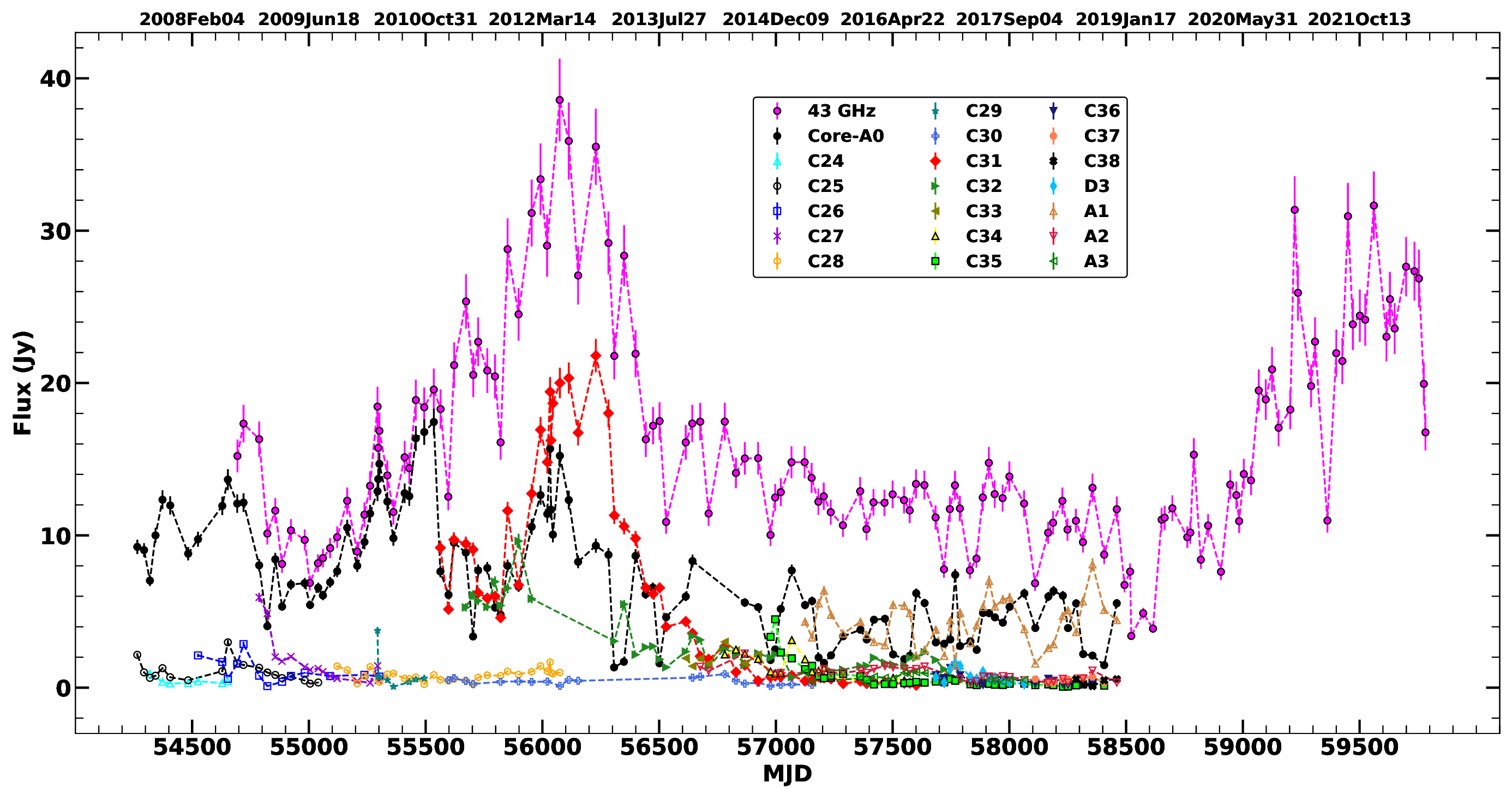}
	\caption{Light curves of individual components of 3C\,279 based on VLBA images at 43 GHz.}
    \label{fig:radio_core_knot_lc}
\end{figure*}

\noindent
A standard explanation of time lags at lower frequencies is in terms of differing opacities. One sees more deeply into the source at higher frequencies --- there is greater synchrotron self-absorption at the lower frequencies --- so changes in emission at higher frequencies are observed earlier, and are less smeared out, than those at lower frequencies.  
Alternatively, even if none of the bands are self-absorbed, the lags and broadening of the flare/outburst profile in the light curve (especially at the lower frequencies) can be understood if the higher-frequency radiation is emitted closer to the site of particle acceleration at, for example, a shock front.
In this case, the lower-frequency emission emerges from more diffuse regions further from the shock, producing the major flux changes, hence variations are delayed and broadened \citep[e.g.,][]{marscher1985,hughes1989ApJ}. 

\subsection{Evolution of Core and Knots}
\begin{figure}
	\centering
 \includegraphics[scale=0.15]{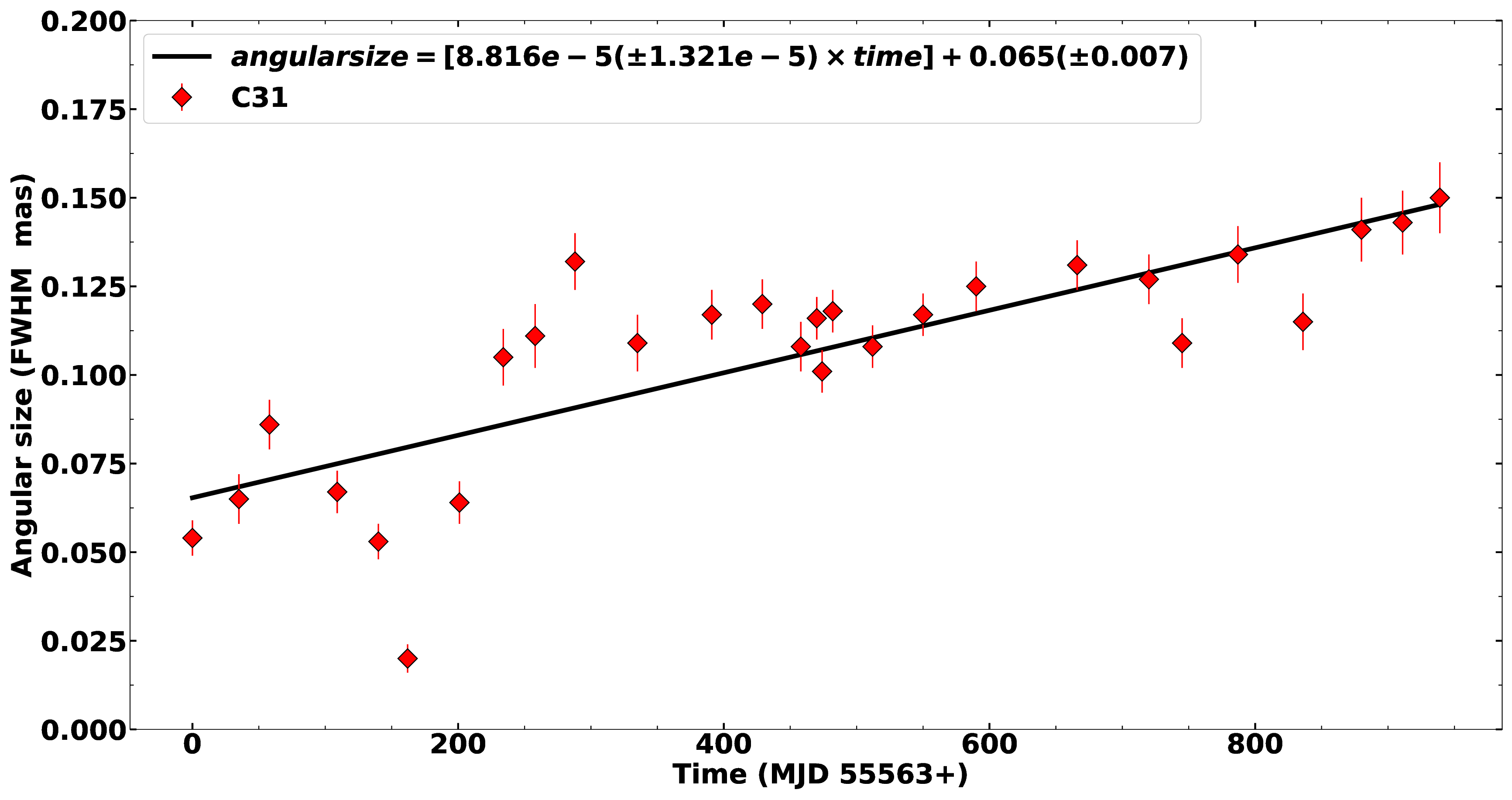}
 \includegraphics[scale=0.15]{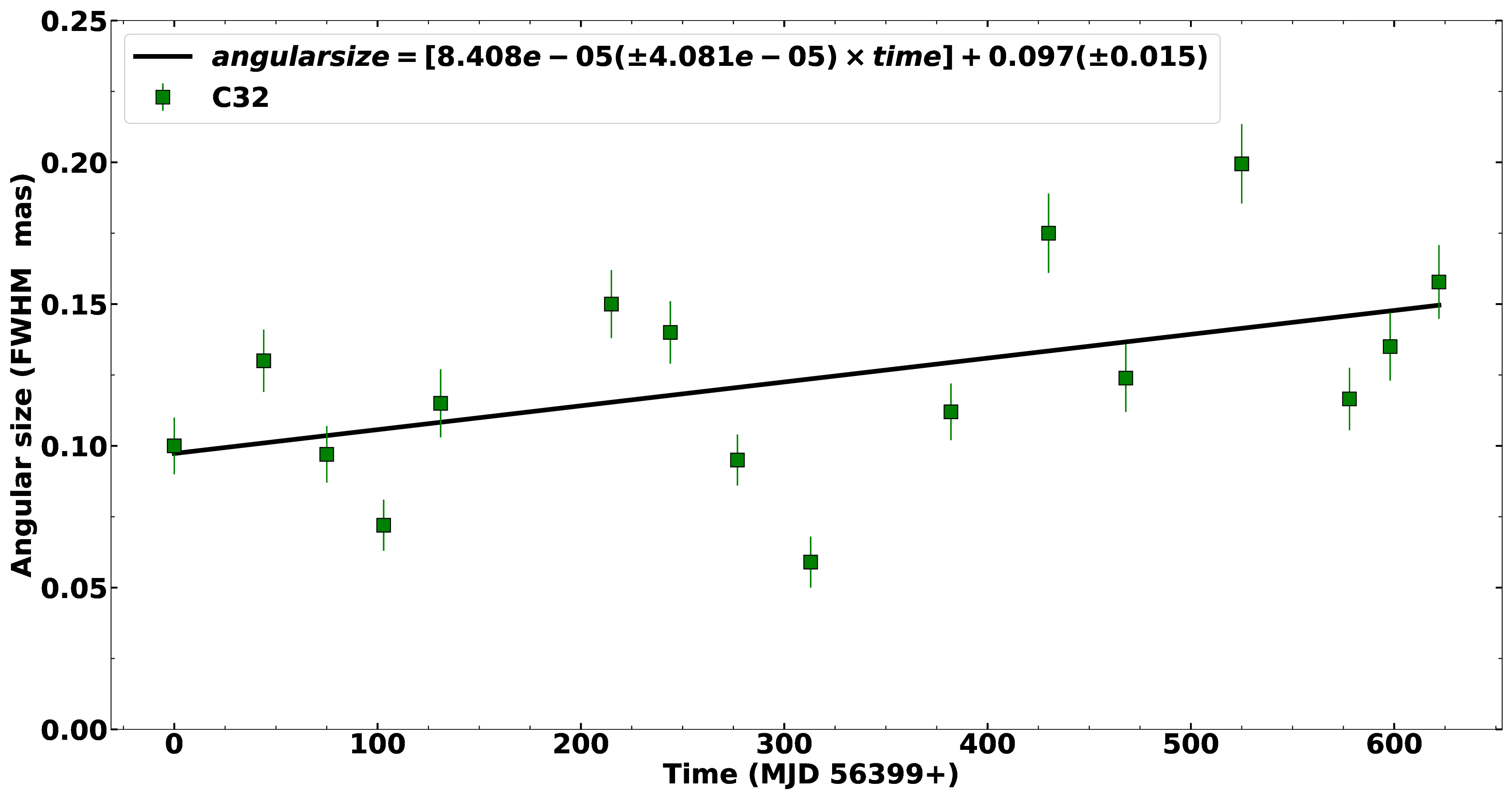}
	\caption{Evolution of the radio knot C31 (top panel) and C32's (bottom panel) angular size vs.\ time along with the straight line fit.}
    \label{fig:knot_C31_C32_size_evolution}
\end{figure}

To understand the relevance of the evolution of the radio core and the individual knots to the observed features in our
work, we have analyzed the results of the VLBA study carried out by \citet{weaver2022ApJS}. These authors presented the kinematics of parsec-scale jets of a sample 
of $\gamma$-ray bright blazars monitored with the VLBA at 43 GHz, covering 
observations from 2007 June to 2018 December.
They found that the jet of 3C\,279 during this time span consisted of the ``core,'' A0, and 17 moving features
(C24, C25, C26, C27, C28, C29, C30 (which includes C30a and C30b), C31, C32, C33,
C34, C35, C36, C37, C38 and D3). Three ``quasi stationary'' features tagged as A1, A2, and A3 are also present. Using the data provided in \citet{weaver2022ApJS}, we have created light curves for the core and individual
knots, as displayed in Figure~\ref{fig:radio_core_knot_lc}. 
From Figure~\ref{fig:radio_core_knot_lc}, we find that knot C31 is main component involved in the 
outburst from MJD $\sim55200$ to $\sim56300$ $-$ a span of about 3 years. (It should be noted that C31 was probably blended with the core A0 during the first part of the outburst.) \\
\\
The flux density of C31 dropped very abruptly (over $\sim100$ days) from MJD $\sim56200$ to $\sim56300$ to end the outburst. This could be due to radiative energy losses, perhaps in combination with adiabatic expansion, which would lower the maximum electron energy as well as the magnetic field, so that the cutoff frequency decreases to $<43$ GHz.
In order to verify that the knots in fact expanded during the outburst, we fit a least-squares straight line to the angular size vs.\ time for C31 and C32 (shown in Fig.~\ref{fig:knot_C31_C32_size_evolution}). Although there are apparently random fluctuations about the 
straight lines, likely owing to the difficulty of measuring angular sizes from the VLBA data, it is apparent that both knots C31 and C32 expanded during the outburst.\\

\noindent
The MJD $\sim55200$ to $\sim56300$ outburst peaked at around the same time at 23, 11, and 8 GHz (Fig.~\ref{fig:multiband_radio_lightcurve}). This could have occurred if the peak was caused by the Doppler factor reaching a maximum and then decreasing with time, so that the fluxes at all frequencies of the flare decreased together. 
At 43 GHz, the peak flux occurred before (Fig.~\ref{fig:multiband_radio_lightcurve}) than at 23, 11, and 8 GHz, 
so the temporal evolution at this higher frequency is due to something else, probably either energy losses (as described above) or an optically thick-to-thin transition.
The position angle vs. time plot \citep[see Fig.~6 of][]{weaver2022ApJS} shows that the direction of knot C31 changes rapidly late in the outburst, so a changing Doppler factor is a reasonable hypothesis. The Doppler beaming factor is given by
\begin{equation}
    \delta = [\Gamma(1-\beta \cos\theta)]^{-1},
\label{Doppler}
\end{equation}
where $\beta$ and $\theta$ respectively are the magnitude (in units of the speed of light) and angle (relative to the line of sight) of the velocity of the knot, and $\Gamma\equiv(1-\beta^2)^{-1/2}$ is the Lorentz factor. The apparent speed in light units in the sky plane, after removal of the redshift dependence, is
\begin{equation}
    \beta_{\rm app} = \beta \sin\theta[1-\beta \cos\theta]^{-1}.
\label{betaapp}
\end{equation}
We can combine these equations to solve for the angle $\theta$:
\begin{equation}
    \theta = \sin^{-1}\Big[{{\beta_{\rm app}}\over{\delta\sqrt{\Gamma^2-1}}}\Big].
    \label{theta}
\end{equation}
The flux density of a knot with a spectral index of 0 (i.e., observed at the spectral peak) depends on the Doppler factor as $F_\nu\propto \delta^3$. The increase in flux density by a factor of $\sim 4$ between MJD 55800 and 56230 could have corresponded to an increase in the Doppler factor by a factor of 1.6. From the data presented in \citet{weaver2022ApJS}, we find that the apparent speed of knot C31 varied from $4.2c$ from MJD 55600 to 55900, to $1.7c$ from MJD 55950 to 56100, to $16c$ after MJD 56200. The second of these intervals included the peak in flux. This, plus the relatively low apparent speed implies that $\theta$ was very close to zero, so that $\delta\sim2\Gamma$ if $\Gamma>>1$. Equation \ref{theta} then becomes $\theta\approx 0.5\beta_{\rm app}\Gamma^{-2}$. We then find that the flux variation can be explained if knot C31 maintained a constant Lorentz factor of
$30\Gamma_{30}$ and the trajectory bent from $\theta=0.13\degr\Gamma_{30}^{-1}$ (MJD 55600--55900) to 
$\theta=0.086\degr\Gamma_{30}^{-1}$ (MJD 55950--56100) and then to $\theta=0.77\degr\Gamma_{30}^{-1}$ (after MJD 56200). We conclude that, because of the high Lorentz factor of the jet flow in 3C\,279, a very slight bending of the trajectory of a knot is sufficient to cause a major outburst after the knot separates from the core. \\
\\

\begin{table*}
  \centering
  \caption{Maximum flux information for the second outburst.}
  \label{tab:max_flux_second_outburst}
\begin{tabular}{cccccc}
\hline
Frequency  & Time period of quiescent flux  & Quiescent flux  & Time of maximum flux  & Maximum flux   & Maximum flux$-$quiescent flux  \\
(GHz)   & (MJD)  & (S$_{\text{quiescent}}$ in Jy) & (MJD) & (S$_{\text{max}}$ in Jy)  & (Jy) \\
\hline
23.0 & 54933.931 $-$ 55339.956  & $12.43\pm0.04$ & 56719.03   & $28.62\pm1.63$  & $16.19\pm1.63$ \\
11.0 & 55072.644 $-$ 55407.813  & $10.23\pm0.03$ & 56713.50   & $31.51\pm0.85$  & $21.27\pm0.85$ \\
~8.0 & 55133.600 $-$ 55373.844  & $9.934\pm0.003$  & 56713.50   & $31.02\pm1.11$  & $21.09\pm1.11$ \\
~4.8 & 55151.259 $-$ 55374.000  & $9.081\pm0.001$  & 56872.75  & $22.64\pm0.23$  & $13.56\pm0.23$  \\
~2.5 & 55255.130 $-$ 55700.845  & $8.71\pm0.01$  & 56992.36  & $15.13\pm0.36$  & ~~
$6.42\pm0.36$  \\
\hline  
  \end{tabular}
 \end{table*}

\begin{figure}
	\centering
 \includegraphics[scale=0.4]{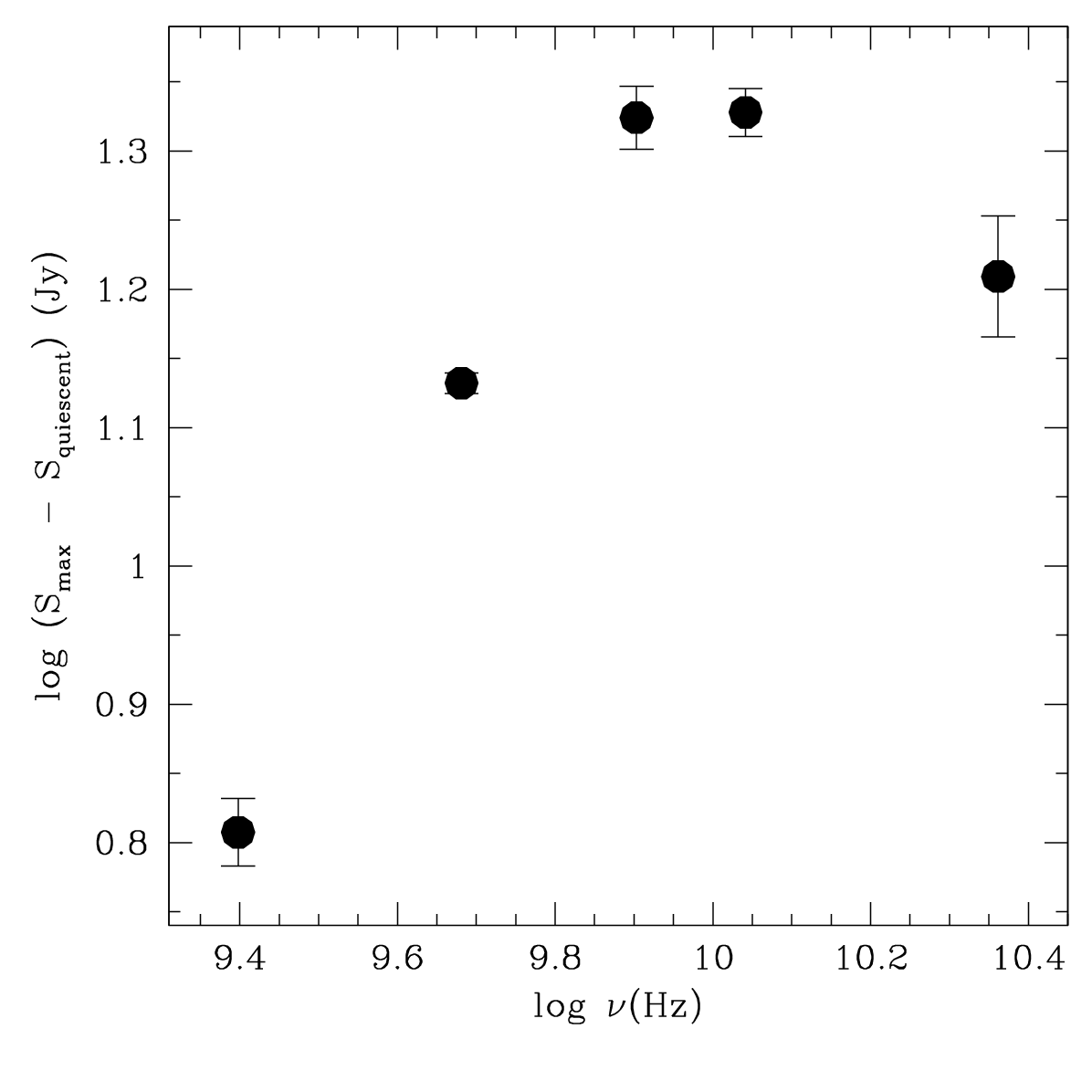}
	\caption{Maximum flux minus quiescent flux vs.\ frequency plot for the second outburst between MJD 56500 and 57000}.
    \label{fig:max_flux_vs_freq_second_outburst}
\end{figure}

\noindent
The outburst from MJD $\sim56500$ to $\sim57000$ was significant only at frequencies $\leq23$ GHz. Hence, it is not apparent on the 43 GHz images.  
From the light curves for the core and individual knots (Fig.\ 4), we noticed that the knots C32 and C33 both appear to be involved during this outburst and exhibited a low-amplitude flare at 43 GHz. To
verify the expansion of the knot during an outburst (as explained above) and to obtain a clearer trend
of such expansion, we only considered the knot C32. This is because the data were more extensive for
knot C32 compared to the sparser observations for knot C33.
Since we have verified that knot C32 expanded during the outburst, an interesting question is whether the adiabatic or synchrotron stage of the expanding shock model of Marscher \& Gear \citep{marscher1985} fits the maximum flux vs.\ frequency dependence of this section of the light curve at frequencies $\leq23$ GHz. To investigate this, we have estimated the peak flux of the outburst at each frequency with the quiescent flux subtracted. The quiescent flux was considered as the average value during the quiescent period before the flux started to rise, i.e., the minimum flux level between MJD $55000$ and $55500$. The details on the time period of consideration for the quiescent flux level and the maximum peak flux of the outburst at different frequencies are provided in Table~\ref{tab:max_flux_second_outburst}. Further, using these values, a plot of maximum flux with subtracted quiescent flux vs.\ frequency is plotted in Figure~\ref{fig:max_flux_vs_freq_second_outburst}. \\

\noindent
The Marscher \& Gear \citep[MG][]{marscher1985} shock-in-jet model predicts a dependence of the maximum flux density $F_m$ observed at frequency $\nu$ during an outburst to be $F_\nu \propto \nu^b$, where $b$ depends on the dominant process by which the radiating electrons lose energy. In order to calculate $b$ from the equations of \citep{marscher1985}, we adopt $s=2.2$ as the slope of
the energy distribution of electrons accelerated at
the shock front \citep[see, e.g.,][]{Sironi2015}, and a magnetic field strength $B$ that depends on radius $r$ as $B\propto r^{-1}$. The shock is assumed to
expand only transverse to the jet initially, but may become a plasmoid that expands longitudinally as well during late stages.
As derived from Table~\ref{tab:max_flux_second_outburst} and displayed in 
Figure~\ref{fig:max_flux_vs_freq_second_outburst}, the value of $b$ changes from $b_1=-0.37\pm0.20$ as the peak flux moves from 23 to 11 GHz, then 
$b_2=0.03\pm0.31$ from 11 to 8 GHz, and 
$b_3=1.15\pm0.11$ from 4.8 to 2.5 GHz.
The prediction of MG is $b=-0.15$ during the synchrotron-loss stage, consistent within the uncertainties with $b_2$ and within $1.1\sigma$ of $b_1$.
The MG value for the later, adiabatic-loss stage is $b=0.51$, which is a considerably weaker dependence than $b_3$.
It is closer to the value $b=0.93$ predicted by the MG adiabatic-loss stage if the expansion is in all three dimensions instead of only laterally.
The early observed value, $b_1$, with the peak flux density increasing as the frequency decreases, is in the same sense as the Compton stage of MG, but much weaker.
However, MG assumed that the seed photons for Compton scattering are from synchrotron radiation within the shock. If, instead, the seed photons are from regions outside the jet (e.g., the broad emission-line region \citep{hayashida2012ApJ}), the Compton losses probably decay more slowly than adopted by MG, with the absolute value of $b$ lower than derived by MG. 
Another possibility is that the Doppler factor increased during the early stages of the outburst, causing $F_m$ to rise more rapidly as the peak frequency decreased than predicted by the MG synchrotron-loss stage.

\section{Conclusions}\label{sec:conclusions}
We have conducted a detailed analysis of quasi-simultaneous multi-band radio flux observations of the blazar 3C\,279. Our data cover the period of 2008 to 2022 at seven different frequencies ranging from 2.5 GHz to 230 GHz. Some of the data were collected from public archives; new observations from various radio telescopes are listed in Table \ref{tab:radio_data_info}. The summary and conclusions from our study are as follows:
\begin{enumerate}
\item[{\bf 1.}] Overall, the multi-band radio light curves exhibit flux variations with prominent outbursts occurring first at higher-frequency and later at lower-frequency bands. Such a trend was further quantified using cross-correlation analysis, indicating that the emission at lower-frequency bands lags that at higher-frequency bands. As is typical for blazars, the flux changes at the higher frequencies lead those at lower frequencies \citep[e.g.,][and references therein]{2001A&A...377..396R,2003A&A...402..151R,2012MNRAS.425.1357G,2014MNRAS.437.3396K,2015A&A...582A.103G}. 
The plots of lag versus frequency are well fit by straight lines with a negative slope, typically $\sim -30$ day GHz$^{-1}$. 
This observational trend possibly 
favours scenarios where time lags at lower frequencies are due to different
opacities, as greater synchrotron self-absorption could occur at the lower frequencies than at the higher ones. Thus, measurements of the light curve time lag in radio bands are a useful way to examine the properties of the opaque apparent base of AGN jets.
Under the shocked jet scenario, frequency-dependent radiative loss timescales could also play a role, especially at high radio frequencies  \citep[e.g.,][]{marscher1985,hughes1989ApJ}.
\\
\item[{\bf 2.}]The multi-band radio light curve exhibits two major outbursts. The first one occurred around MJD $\sim55200$ to $\sim56300$. The second one, which was significant only at frequencies $\leq23$ GHz, took place from MJD $\sim56500$ to $\sim57000$. From the radio knots identified in 43 GHz VLBA images, we noticed that knot C31 was found to be mainly involved in the first outburst and radio knots C32 \& C33 were involved during the second outburst. 
The observed abrupt drop in flux density of knot C31 over $\sim100$ days during the first outburst could have been caused by radiative energy losses in combination with the adiabatic expansion of the emitting knot. Such expansion of a knot's size with time was also found for component C32.
 \\
\item[{\bf 3.}] We conclude that variation in the Doppler factor, along with very slight bending of the trajectory of a knot, is a reasonable hypothesis for the cause of the major outburst after knot C31 detached from the core in the VLBA images. 
\\
\item[{\bf 4.}] The dependence of maximum flux vs.\ frequency was considered so as to compare it with the adiabatic or synchrotron-loss stages of the MG expanding shock model \citep{marscher1985} during the second outburst. It was found that the MG synchrotron-loss stage is consistent with the data in the (23 $-$ 11) GHz and (11 $-$ 8) GHz frequency ranges. During the latter part of the outburst, the observed dependence was stronger than that predicted by the MG adiabatic-loss stage if the knot only expanded in the direction transverse to the jet axis. However, if the expansion during this late stage was in all three dimensions, the MG prediction agrees with the observed behavior within the uncertainties. 
\\
\item[{\bf 5.}] This study demonstrates the importance of long-term radio monitoring programs of blazars, over a wide range of frequency bands, toward a better understanding of the physical processes that occur in their relativistic jets.
Further long-term multi-band studies of objects belonging to different blazar sub-classes have great potential to test different physical scenarios that have been proposed to explain their behavior.
\end{enumerate}

\section*{Acknowledgements}
The observations were carried out with the RATAN-600 radio telescope and funded by the Ministry of Science and Higher Education of the Russian Federation under contract 075-15-2022-1227. The research at UMRAO was funded in part by a series of grants from the NSF (most recently AST-0607523) and by a series of Fermi G.I. awards from NASA (NNX09AU16G, NNX10AP16G, NNX11AO13G, and NNX13AP18G). Funds for the operation of UMRAO were provided by the University of Michigan. This study makes use of VLBA data from the VLBA-BU Blazar Monitoring Program (BEAM-ME and VLBA-BU-BLAZAR; \url{http://www.bu.edu/blazars/BEAM-ME.html}), funded by NASA through the Fermi Guest Investigator Program. The VLBA is an instrument of the National Radio Astronomy Observatory. The National Radio Astronomy Observatory is a facility of the National Science Foundation operated by Associated Universities, Inc. The XAO-NSRT is operated by the Urumqi Nanshan Astronomy and Deep Space Exploration Observation and Research Station of Xinjiang (XJYWZ2303). The Submillimeter Array (SMA) is a joint project between the Smithsonian Astrophysical Observatory and the Academia Sinica Institute of Astronomy and Astrophysics and is funded by the Smithsonian Institution and the Academia Sinica. We recognize that Maunakea is a culturally important site for the indigenous Hawaiian people; we are privileged to study the cosmos from its summit. F-GAMMA utilised several facilities in cm, mm, sub-mm, infrared and optical bands, achieving an unprecedented coverage for the study of the spectral evolution of powerful relativistic jets in AGNs. The pivoting facilities were the 100 m radio telescope in Effelsberg (Germany), the 30 m IRAM radio telescope at Pico Veleta (Spain) and the 12-m APEX telescope (Chile). The frequency coverage included 12 frequencies between 2.64 and 345 GHz. \\
\\
We thank the anonymous reviewer for useful comments which helped us to improve the manuscript.
We are thankful to E. Ros for carefully reading the manuscript and for providing useful comments. KMA acknowledges funding from an ARIES Regular Post-Doctoral Fellowship (AO/RA/2022/1960). ACG is partially supported by Chinese Academy of Sciences (CAS) President’s International Fellowship Initiative (PIFI) (grant no. 2016VMB073). YS and TM are supported in the framework of the national project ‘Science’ by the Ministry of Science and Higher Education of the Russian Federation under the contract 075-15-2020-778. LC and XL's work is supported by the Xinjiang Regional Collaborative Innovation Project (2022E01013), the National Science Foundation of China (12173078 and 11773062) and the Chinese Academy of Sciences (CAS) `Light of West China' Program (2021-XBQNXZ-005). XL is also supported by the National Key R$\&$D Intergovernmental Cooperation Program of China (2023YFE0102300). YYK was supported by the M2FINDERS project which has received funding from the European Research Council (ERC) under the European Union’s Horizon 2020 Research and Innovation Programme (grant agreement No. 101018682). 

\section*{Data Availability}
This work has made use of radio fluxes published by \citet{angelakis2019}.
UMRAO radio data are available at \url{https://dept.astro.lsa.umich.edu/datasets/umrao.php}, 
VLBA-BU Blazar Monitoring Program (BEAM-ME and VLBA-BU-BLAZAR) data are available at 
\url{https://www.bu.edu/blazars/VLBA_GLAST/3c279.html} and SMA data are available at 
\url{http://sma1.sma.hawaii.edu/callist/callist.html}.
The RATAN-600 data underlying this article are available in the BLcat online catalogue at \url{https://www.sao.ru/blcat/} \citep{mingaliev2014,sotnikova2022AstBu}
or from the corresponding author upon reasonable request. XAO data can be obtained from LC (cuilang@xao.ac.cn).



\bibliographystyle{mnras}
\bibliography{references} 








\bsp	
\label{lastpage}
\end{document}